\documentclass[aps,prd,superscriptaddress]{revtex4}

\usepackage{amsfonts}

\usepackage{amsmath}
\usepackage{graphicx}
\usepackage{color}
\usepackage{mathtools,leftidx}

\usepackage{soul}

\DeclareMathAlphabet\mathbfcal{OMS}{cmsy}{b}{n}

\begin{document}

\title{Hamiltonian analysis of ModMax nonlinear electrodynamics in the first order formalism}

\author{C. A. Escobar}
\email{carlos.escobar@ciencias.unam.mx}
\affiliation{Departamento  de  F\'isica,  Universidad  Aut\'onoma  Metropolitana-Iztapalapa, San Rafael Atlixco 186, 09340 Ciudad de M\'{e}xico, M\'{e}xico}

\author{Rom\'an Linares}
\email{lirr@xanum.uam.mx}
\affiliation{Departamento  de  F\'isica,  Universidad  Aut\'onoma  Metropolitana-Iztapalapa, San Rafael Atlixco 186, 09340 Ciudad de M\'{e}xico, M\'{e}xico}

\author{B. Tlatelpa-Mascote}
\email{brendatlatelpa19@gmail.com }
\affiliation{Departamento  de  F\'isica,  Universidad  Aut\'onoma  Metropolitana-Iztapalapa, San Rafael Atlixco 186, 09340 Ciudad de M\'{e}xico, M\'{e}xico}

\begin{abstract}
In this work we study the so-called ModMax nonlinear electrodynamics, which is a novel model designed to preserve duality rotations and conformal transformations, such as the Maxwell's equations do. This model allows to study diverse gravitational phenomena when is coupled to General Relativity, in particular charged black holes and gravitational waves. In the present work we focus in the dynamics and Hamiltonian analysis of the model. Specifically, we analyze the propagation of the discontinuities of the field and obtain the corresponding dispersion relations. To perform the Hamiltonian analysis we adopt the first order formalism develop by Pleba\'nski and follow the Dirac method for theories with constraints. We derive the effective Hamiltonian, classify all the constraints and identify the degrees of freedom. We prove that the effective Hamiltonian is strictly bounded from below and investigate the existence of non trivial minima.      
\end{abstract}

\maketitle

\section{Introduction}

Studies on nonlinear electrodynamics (NLED) have a long history in physics and dates back from 1934 when Born and Infeld \cite{BI1} proposed a nonlinear modification of Maxwell theory with the aim to correct the infinities of the self-energy and the fields of a point charge. Such model is constructed from the only Lorentz electromagnetic invariants $F=F_{\mu\nu}F^{\mu\nu}$ and $G=F_{\mu\nu}\tilde{F}^{\mu\nu}$ \cite{Urru} and it has been shown that it can be derived as an effective action from quantized string theory \cite{ST1}. As a second well known nonlinear electrodynamics example we found the Heisenberg and Euler model \cite{HE1}, which is capable of explaining scattering processes between photons and creation of electron-positron pairs. In a different context, in recent decades have emerged some NLED models to explain the accelerate expansion of the universe \cite{AC1} and galactic magnetic fields \cite{GM1}. Solutions in curved space backgrounds (black holes) have been also extensively studied in the framework of nonlinear electrodynamics \cite{BL1,BL2}. Last but not least, we found the arcsin \cite{SI12}, the Maxwell Power Law \cite{MP1}, the Logarithmic \cite{LL1}, the rational \cite{RL1} and the exponential \cite{EX1} nonlinear electrodynamics models.

Recently, it has been proposed a NLED model that preserves both $SO(2)$ electromagnetic duality invariance and conformal invariance. It is called the ModMax model \cite{MM1} and, despite it not long since came up, a plenty of studies have been carried out when is coupled to General Relativity. It is worth to mention that the first NLED model which is duality and conformal invariant was obtained by Bialynicki-Birula as the strong field limit of the Hamiltonian formulation of the Born-Infeld theory in \cite{BB2}, and whose properties were later discussed in \cite{BB3}. A distinguished characteristic of the ModMax model is that is the only one which has Maxwell's theory in the zero limit of its coupling constant.  A simple derivation of the theory's ModMax Lagrangian is available in \cite{14}. In Ref. \cite{Roman1} charged black holes solutions as well as gravitational waves were analyzed. In the resulting  black hole geometry, it was shown that the relevant $\gamma$-ModMax parameter acts as a screening factor allowing configurations where the mass of the black hole is smaller than its charge. A thermal stability study of the black holes is given in \cite{16}. Studies on Taub-NUT spaces \cite{17} to construct AdS wormholes and some thermodynamic features are analyzed in Ref. \cite{Roman2}. Conformal scalar NUT-like dyons have been also constructed \cite{19}.

Just like standard Maxwell theory, ModMax possesses an interesting topology structure, in particular exhibits hopfion-Ra\~{n}ada null knots, which correspond to topologically nontrivial configurations of electric and magnetic fields \cite{Knots}. 
The prescription for $\mathcal{N}=1$ supersymmetrization of ModMax is developed in Ref. \cite{SSS1}, where is also proved that its coupling to supergravity results super-Weyl invariant. An alternative derivation of the ModMax model in the context of superconformal duality-invariant models was presented in Ref. \cite{BB3a}. Also, the $U(1)$ duality-invariant conformal higher-spin generalization of ModMax, bosonic case and $\mathcal{N}=1$ superconformal case, has been formulated in Ref. \cite{HS1a}.

In the present work we are interested in both the dynamics and Hamiltonian analysis of the ModMax model. Specifically, we study the propagation of linear shock-waves associated with the discontinuities of the field in the limit of geometrical optics and perform the Hamiltonian analysis in detail. We follow the Dirac method \cite{Dirac1} for theories with constraints and analyze the conditions over the effective Hamiltonian to be bounded from below and the existence of one or more local minima at nonzero field values. Given the nonlinear nature of the ModMax model, the first order formalism developed by Pleba\'nski \cite{Ple1} is employed. 

The paper is organized as follows: In section \ref{Intro} we present the ModMax model and review some of the main properties, such as equations of motion and energy-momentum tensor. Section \ref{WP1} is devoted to  study the propagation of the discontinuities of the field. The Hamiltonian analysis and constraints, as well as the conditions for a positive effective Hamiltonian, is developed in section \ref{HA1}. We finalize in section \ref{CON1} summarizing  our results and giving further concluding remarks. Throughout the paper, natural units are assumed $\hbar=c=1$ and the metric convention ($+, -, -, -$) is followed.

\section{ModMax Electrodynamics}
\label{Intro}

In this section we present the ModMax nonlinear electrodynamics. The Lagrangian density is given by 

\begin{equation}
\mathcal{L}_{\textrm{ModMax}}(x,y)=-x \cosh\gamma+(x^2+y^2)^{\frac{1}{2}} \sinh\gamma,
\label{L1}
\end{equation}
 
where $\gamma$ is a dimensionless real parameter and 

\begin{equation}
x:=\frac{1}{4}F^{\mu\nu}F_{\mu\nu}, \quad\quad y:= \frac{1}{4}F^{\mu\nu}\tilde{F}_{\mu\nu}
\end{equation}

correspond to the standard Lorentz invariants of the electromagnetic field, being

\begin{equation}
\tilde{F}^{\mu\nu}:=\frac{1}{2}\epsilon^{\mu\nu\alpha\beta}F_{\alpha\beta}
\end{equation}

the dual of $F^{\mu\nu}$. Here we use the convention $\epsilon^{0123} = -\epsilon_{0123} = 1$ for the Levi-Civita symbol. The standard Maxwell electrodynamics is recovered by taking $\gamma=0$. As we will see later, a carefully interpretation of the limit $\gamma\rightarrow0$ should be done due that as long as $\gamma\neq0$, even at $\gamma\ll1$, the ModMax model presents specific features, such as birefringence. Furthermore, restrictions on $\gamma$ should be imposed. In particular, in Ref. \cite{MM1} it has been proved that causality and unitary conditions lead to the relation $\gamma>0$.

The Lagrange equations of motion read as follows

\begin{equation}
(\mathcal{L}_x F^{\mu\nu}+\mathcal{L}_y \tilde{F}^{\mu\nu}),_\nu=0,
\end{equation}

where we have defined

\begin{equation}
\mathcal{L}_x:=\frac{\partial \mathcal{L}_{\textrm{ModMax}}}{\partial x}; \quad\quad \mathcal{L}_y:=\frac{\partial \mathcal{L}_{\textrm{ModMax}}}{\partial y}.
\end{equation}

The energy-momentum tensor is given by

\begin{equation}
T^{\mu\nu}=(F^{\mu\sigma}F^{\nu}\,_{\sigma}-x\eta^{\mu\nu})\bigg(\cosh\gamma-\frac{x}{\sqrt{x^2+y^2}}\sinh\gamma\bigg).
\end{equation}

As is expected $T^{\mu}\,_\mu=0$ as a consequence of the conformal invariance of the theory.

\section{Wave propagation}
\label{WP1}

Propagation in nonlinear electrodynamics is far from a trivial task. Several approaches have been developed to analyze this problem, in particular those due to Boillat \cite{Boillat1}, Bialynicka-Birula and Bialynicki-Birula \cite{Birula1} and more recently by Novello et al. \cite{NovelloSalim}. These approaches have shown that propagation of shock waves is an inherent peculiarity for most of the models of nonlinear electrodynamics, with unique exceptions  such as the Born-Infeld model \cite{BornN}. In the limit of geometrical optics, the propagation of linear shock waves can be analyzed by means of the method of the discontinuities of the field proposed by Hadamard \cite{Hadamard1}. It consists in considering a surface of discontinuity of the field $\Sigma$ defined by

\begin{equation}
z(x^\mu)=0.
\end{equation}

such that it divides the spacetime in two different regions $U^-$ and $U^+$ ($z<0$ and $z>0$, respectively). For any function on the spacetime $f(x^{\mu})$, it is defined its discontinuity on $\Sigma$ as

\begin{equation}
[f(x^\mu)]_{\Sigma} := \lim_{\{P^\pm\}\rightarrow P} [f(P^+)-f(P^-)]
\end{equation}
where  $P^+$, $P^-$ and $P$ belong to $U^+$, $U^-$ and $\Sigma$ respectively. For the field strength tensor $F^{\mu\nu}$ this method produces

\begin{equation}
[F_{\mu\nu}]_\Sigma=0, \quad\quad [F_{\mu\nu},_\lambda]=f_{\mu\nu}k_\lambda,
\end{equation}  

where $f_{\mu\nu}$ corresponds to the discontinuities of the field on the surface $\Sigma$ and $k_\lambda$ represents the wave propagation four vector. Following Ref. \cite{NovelloSalim} and using the Bianchi identities, we found that the propagation of the field discontinuities is described by two different polarization modes satisfying  

\begin{equation}
k^2_+=0; \quad\quad k^2_-=\bigg(\frac{\sinh\gamma(x\sinh\gamma-\cosh\gamma\sqrt{x^2+y^2})}{x^2+y^2+y^2\sinh\gamma}\bigg) F^{\lambda\mu}F^{\nu}\,_\lambda (k_-)_\mu (k_-)_\nu
\label{ks1}
\end{equation}

which indicates birefringence. Notice that in the limit $\gamma\rightarrow0$, $k^2_-=0$, as it should be. In the general case, the light cone structure, which dictates the photon paths, is given by the condition 

\begin{equation}
g^{\mu\nu}k_\mu k_\nu=0,
\end{equation}
where the $k_\mu$ are null vectors in the effective geometry but are not null in the background geometry.  The effective geometry is generated by the self-interaction of the field with itself due the nonlinear character of the theory. Rewriting Eq. (\ref{ks1}) we arrive to the effective metrics

\begin{equation}
g^{\mu\nu}_+=\eta^{\mu\nu},\quad\quad g^{\mu\nu}_-=\eta^{\mu\nu}\bigg(\frac{\textrm{sech}\,\gamma(x^2+y^2\cosh\gamma)}{x^2+y^2}\bigg)+\frac{\sinh\gamma(\sqrt{x^2+y^2}-x\tanh\gamma)}{x^2+y^2} F^{\lambda\mu}F^{\nu}\,_\lambda.
\end{equation}

Once again, in the limit $\gamma\rightarrow0$, we obtain $g^{\mu\nu}_+=g^{\mu\nu}_-=\eta^{\mu\nu}$. An analysis and classification of effective metrics can be found in Ref.\cite{EffeRef}.

\section{Hamiltonian Analysis}
\label{HA1}

In this section we delve into the Hamiltonian analysis of ModMax theory; in particular, we focus on the constraints structure and the effective Hamiltonian. Given the nonlinear character of the model it is more convenient to adopt the first order formalism developed by Pleba\'nski for theories of nonlinear electrodynamics.  To do so, it is introduced the antisymmetric tensor $P^{\mu\nu}$ through the Legendre transformation

\begin{eqnarray}
\label{Lag1}
\hat{\mathcal{L}}_{\textrm{ModMax}}(P_{\mu\nu},A_\mu) :&=&-\frac{1}{2}P^{\mu\nu}F_{\mu\nu}-\hat{V}, \\ \nonumber
&=&- P^{\mu\nu}\partial_\mu A_\nu-\hat{V}.
\end{eqnarray}

where $\hat{V}(P_{\mu\nu},A_\mu)$ corresponds to the ModMax Lagrange density expressed in terms of the variables $P_{\mu\nu},A_\mu$, i.e. $\hat{V}=\mathcal{L}_{\textrm{ModMax}}(P_{\mu\nu},A_\mu)$. Defining the invariants

\begin{equation}
\tilde{x}:=\frac{1}{4}P^{\mu\nu}P_{\mu\nu}, \quad\quad \tilde{y}:= \frac{1}{4}P^{\mu\nu}\tilde{P}_{\mu\nu},
\end{equation}

and treating the vector potential $A_\mu$ and the antisymmetric tensor $P^{\mu\nu}$ as independent fields, the equations of motion read as follows  

\begin{eqnarray}
\partial_\nu P^{\mu\nu}&=&0, \label{E12a} \\  F_{\mu\nu}&=& -\hat{V}_{\tilde{x}} P_{\mu\nu}-\hat{V}_{\tilde{y}} \tilde{P}_{\mu\nu} 
\label{Fmunu}.
\end{eqnarray}

where we have defined 

\begin{equation}
\hat{V}_{\tilde{x}}= \frac{\partial \hat{V}}{\partial \tilde{x}},\quad\quad  \hat{V}_{\tilde{y}}= \frac{\partial \hat{V}}{\partial \tilde{y}}.
\end{equation}

From the definition of $F_{\mu\nu}:= \partial_\mu A_\nu-\partial_\nu A_\mu$, we  also find the Bianchi identities 

\begin{equation}
\partial_\mu \tilde{F}^{\mu\nu}=\frac{1}{2}\partial_\mu \epsilon^{\mu\nu\rho\sigma}F_{\rho\sigma}=0.
\label{Bianchi1}
\end{equation}  

Notice that Eq. (\ref{Fmunu}) corresponds to the constitutive relation and allows us to express $\mathcal{L}_{\textrm{ModMax}}(x,y)$ in terms of the invariants $\tilde{x}$ and $\tilde{y}$. To do so, we take the variation of the action $S=\int\mathcal{L}_{\textrm{ModMax}}(x,y) d^4x$ with respect to $F^{\mu\nu}$ and comparing it with the variation of the action in Eq. (\ref{Lag1}) it follows that

\begin{eqnarray}
P_{\mu\nu}&=&- \mathcal{\hat{L}}_{x} F_{\mu\nu}-\mathcal{\hat{L}}_{y} \tilde{F}_{\mu\nu},
\label{P1a}
\end{eqnarray}

which expresses the inverse of the constitutive relation (\ref{Fmunu}). An straightforward calculation shows that

\begin{equation}
\hat{V}(\tilde{x},\tilde{y})=\mathcal{L}_{\textrm{ModMax}}(\tilde{x},\tilde{y})=-\tilde{x} \cosh\gamma-(\tilde{x}^2+\tilde{y}^2)^{\frac{1}{2}} \sinh\gamma
\end{equation}

Notice that we can obtain this result from Eq. (\ref{L1}) with the replacement $\mathcal{L}_{\textrm{ModMax}}\rightarrow-\hat V$ together $x\rightarrow-\tilde{x}$. This transformation is associated with the fact that, by construction of the first order formalism, the model preserves its invariance under duality transformations.

The equations (\ref{E12a}) and (\ref{Bianchi1}) can be cast into the form of the standard Maxwell equations in material media by defining the vectors $\vec{D}$, $\vec{E}$, $\vec{H}$ and $\vec{B}$ as follows

\begin{equation}
P^{\mu\nu}=\left(
\begin{matrix}
0 & -D_x & -D_y & -D_z\\
D_x & 0 & -H_z & H_y\\
D_y & H_z & 0 & -H_x \\
D_z & -H_y & H_x & 0 \\
\end{matrix}
\right)
\label{PP1}
\end{equation}

and

\begin{equation}
F^{\mu\nu}=\left(
\begin{matrix}
0 & -E_x & -E_y & -E_z\\
E_x & 0 & -B_z & B_y\\
E_y & B_z & 0 & -B_x \\
E_z & -B_y & B_x & 0 \\
\end{matrix}
\right)
\end{equation}

The invariants $x$, $y$, $\tilde{x}$ and $\tilde{y}$ are written as

\begin{eqnarray}
x&=&\frac{1}{2}(\vec{B}^2-\vec{E}^2), \quad\quad y=-\vec{B}\cdot \vec{E} \\
\tilde{x}&=&\frac{1}{2}(\vec{H}^2-\vec{D}^2), \quad\quad \tilde{y}=-\vec{H}\cdot \vec{D}. \\
\end{eqnarray}

The constitutive relations (\ref{Fmunu}) and (\ref{P1a}) take the form

\begin{equation}
\left(
\begin{matrix}
\vec{E} \\
\vec{B} \\
\end{matrix}
\right)
=\left(
\begin{matrix}
-\hat{V}_{\tilde{x}} & -\hat{V}_{\tilde{y}} \\
\hat{V}_{\tilde{y}} & -\hat{V}_{\tilde{x}} \\
\end{matrix}
\right)\left(
\begin{matrix}
\vec{D} \\
\vec{H} \\
\end{matrix}
\right)
\end{equation}

and

\begin{equation}
\left(
\begin{matrix}
\vec{D} \\
\vec{H} \\
\end{matrix}
\right)
=\left(
\begin{matrix}
-\hat{\mathcal{L}}_{\tilde{x}} & -\hat{\mathcal{L}}_{\tilde{y}} \\
\hat{\mathcal{L}}_{\tilde{y}} & -\hat{\mathcal{L}}_{\tilde{x}} \\
\end{matrix}
\right)\left(
\begin{matrix}
\vec{E} \\
\vec{B} \\
\end{matrix}
\right)
\end{equation}

Finally the equations of motion can be written as

\begin{eqnarray}
\vec{\nabla}\cdot \vec{D}&=&0, \\
\vec{\nabla}\times\vec{H}-\frac{\partial \vec{D}}{\partial t}&=&0, \\
\vec{\nabla}\cdot \vec{B}&=&0, \\
\vec{\nabla}\times\vec{E}+\frac{\partial \vec{B}}{\partial t}&=&0. \\
\end{eqnarray}

To perform the Hamiltonian  analysis, following the Dirac method \cite{Dirac1},  we write the Lagrange density in Eq. (\ref{Lag1}) as follows

\begin{equation}
\hat{\mathcal{L}}_{\textrm{ModMax}}(P_{\mu\nu},A_\mu)=A_i \partial _0 P^{0i}-(\partial_i A_\nu)P^{i\nu}- \hat{V}(\tilde{x},\tilde{y}).
\end{equation}

Dropping the subindex ``ModMax'' in $\hat{\mathcal{L}}_{\textrm{ModMax}}$, the canonical momenta 

\begin{equation}
\Pi_0^A=\frac{\partial \hat{\mathcal{L}}}{\partial \dot{A}_0}, \quad \Pi_i^A=\frac{\partial  \hat{\mathcal{L}}}{\partial \dot{A}^i}, \quad \pi_i=\frac{\partial  \hat{\mathcal{L}}}{\partial \dot{P}^{0i}}, \quad \pi_{ij}=\frac{\partial  \hat{\mathcal{L}}}{\partial \dot{P}^{ij}}
\end{equation}

which obey the canonical algebra

\begin{eqnarray}
&&\{A_0(\vec{x}), \Pi_0^A(\vec{y})\}=\delta(\vec{x}-\vec{y}), \quad\quad\quad \{A_i(\vec{x}), \Pi^{Aj}(\vec{y})\}=\delta_i\,^j\delta(\vec{x}-\vec{y}), \\ \nonumber
&&\{P_{0i}(\vec{x}), \pi^{j}(\vec{y})\}=\delta_i\,^j\delta(\vec{x}-\vec{y}),\quad\quad 
\{P_{ij}(\vec{x}), \pi^{lm}(\vec{y})\}=(\delta_i\,^l\delta_j\,^m-\delta_i\,^m\delta_j\,^l)\delta(\vec{x}-\vec{y}),
\end{eqnarray}

lead to the set of primary constraints  

\begin{equation}
\Phi^1=\Pi_0^A\approx0,\quad\quad \Phi_i^2=\Pi_i^A\approx0, \quad\quad \Phi_i^3=\pi_i-A_i\approx0,\quad\quad \Phi_{ij}^4=\pi_{ij}\approx0.
\end{equation}

As usual, the symbol $\approx$ means that the constraints vanish weakly (in Dirac's terminology), i. e. they are zero on the constraint surface. The canonical Hamiltonian results

\begin{equation}
\mathcal{H}_0=(\partial_i A_0)P^{i0}+(\partial_i A_j)P^{ij}+\hat{V}(\tilde{x},\tilde{y}).
\end{equation}

According to the Dirac's method, the next step is to consider the extended Hamiltonian,

\begin{eqnarray}
\mathcal{H}_E= \mathcal{H}_0 + \lambda^1 \Phi^1+\lambda_i^2 \Phi^{2i}+\lambda_i^3\Phi^{3i}+\lambda_{ij}^4 \Phi^{4ij}
\end{eqnarray}

being $\lambda^1, \lambda_i^2, \lambda_i^3$ and $\lambda_{ij}^4$ Lagrange multipliers, and study the time evolution of the constraints $\dot{\Phi}^k=\{\Phi^k,\mathcal{H}_E\}$ to ensure that they are preserved all the time. Imposing that $\dot{\Phi}^k$ vanish weakly, we arrive to the conditions

\begin{eqnarray}
&&\Delta^1=\partial_i P^{i0}\approx0, \\ \nonumber  
&&\Delta^2_i=\lambda_i^3-\partial_m P^{mi}\approx0,\\ \nonumber
&&\Delta_i^3=\partial^i A_0+\hat{V}_{\tilde{x}} P^{0i}+\hat{V}_{\tilde{y}}\tilde{P}^{0i}-\lambda_i^2\approx0 \\ \nonumber
&&\Delta_{ij}^4=\partial_{i}A_j-\partial_j A_i+\hat{V}_{\tilde{x}}P_{ij}+\hat{V}_{\tilde{y}}\tilde{P}_{ij}\approx0,
\end{eqnarray}

The Lagrange multipliers $\lambda_i^3$ and $\lambda_i^2$ are fixed by the conditions $\Delta^2_i\approx0$ and $\Delta^3_i\approx0$, respectively. An straightforward calculation shows that the time evolution of $\Delta^1$ and $\Delta_{ij}^4$ does not produce additional constraints. The Dirac's method stops and we need to classify the constraints $\{\Phi^1,\Phi^2_i,\Phi^3_i,\Phi^4_{ij},\Delta^1,\Delta^4_{ij}\}$ into first-class and second-class. Let us define a new set of constraints as follows

\begin{eqnarray}
&&\Theta^1\doteq\Phi^1=\Pi_0^A\approx0, \\
&&\Theta^2\doteq \Delta_1+\partial^i \Phi_i^2=\partial_i P^{i0}+\partial^i\Pi_i^A\approx0, \\
&&\Theta^3_i\doteq \Phi_i^2=\Pi_i^A\approx0, \\
&&\Theta^4_i\doteq \Phi_i^3=\pi_i-A_i\approx0, \\
&&\Theta^5_{ij}\doteq\Phi^4_{ij}=\pi_{ij}\approx0, \\
&&\Theta^6_{ij}\doteq\Delta^4_{ij}=\partial_{i}A_j-\partial_j A_i+\hat{V}_{\tilde{x}}P_{ij}+\hat{V}_{\tilde{y}}\tilde{P}_{ij}\approx0.
\end{eqnarray} 

Evaluating the Poisson brackets between the constraints $\Theta^a$, $a=1,2,..,6$, it can be proved that $\Theta^1$ and $\Theta^2$ are first-class constraints. The determinant of the matrix of Poisson brackets of the remaining constraints $\{\Theta^3_i,\Theta^4_i,\Theta^5_{ij},\Theta^6_{ij}\}$ satisfy

\begin{eqnarray}
|\{\Theta^a(\vec{x}),\Theta^b(\vec{y}) \}|=\hat{V}_{\tilde{x}}^2 S^2 \delta^3(\vec{x}-\vec{y})
\end{eqnarray}

where 

\begin{equation}
S=\frac{(\tilde{x}\sinh\gamma-\cosh\gamma\sqrt{\tilde{x}^2+\tilde{y}^2})(\sinh\gamma(\tilde{x}^2(D^2+\tilde{x})+\tilde{y}^2(H^2+3\tilde{x}))-\cosh\gamma(\tilde{x}^2+\tilde{y}^2)^\frac{3}{2})}{(\tilde{x}^2+\tilde{y}^2)^2}
\end{equation}

with

\begin{equation}
D^2=\vec{D}\cdot\vec{D},\quad\quad H^2=\vec{H}\cdot\vec{H}.
\end{equation}

As long as $S\neq0$, the constraints $\{\Theta^3_i,\Theta^4_i,\Theta^5_{ij},\Theta^6_{ij}\}$ form a set of second-class constraints. Notice that the condition $(\tilde{x}^2+\tilde{y}^2)=0$ it may lead to a singular behavior and it can be seen as a jump between the ModMax model and standard Maxwell electrodynamics, see Eq. (\ref{L1}). Given that the model contains $10$ variables in the coordinate space ($4$ in $A^\mu$ and $6$ in $P^{\mu\nu}$), plus $2$ first-class constrains and $12$ second-class constrains, the counting of degrees of freedom (d.o.f.) tell us that the model has

\begin{equation}
\textrm{\#d.o.f.}= 2\times10-2\times 2-12= 4
\end{equation} 

in phase space.

To construct the effective Hamiltonian, we first fix the gauge degrees of freedom associated with the first-class constraints $\Theta^1$ and $\Theta^2$ imposing the gauge-fixing constraints 

\begin{equation}
\Sigma^1=A_0,\quad\quad \Sigma^2=\partial_i A^i.
\end{equation}

The set $\{\Theta^1,\Theta^2,\Sigma^1,\Sigma^2\}$ is second-class and now we can proceed to analyze the physical degrees of freedom of the model. The effective Hamiltonian, which is obtained from $\mathcal{H}_{E}$ by setting all the second-class constraints strongly equal to zero, reduces to

\begin{equation}
\mathcal{H}_{eff}= \partial_i A_j P^{ij}+\hat{V}(\tilde{x},\tilde{y}),
\end{equation}

which, using the constraint $\Theta^6_{ij}=0$, can be rewritten as

\begin{eqnarray}
\mathcal{H}_{eff}&=& -\frac{1}{2}(\hat{V}_{\tilde{x}} P_{ij}P^{ij}+\hat{V}_{\tilde{y}} P_{ij}\tilde{P}^{ij}+\hat{V}(\tilde{x},\tilde{y}) \\ \nonumber
&=&-H^2\hat{V}_{\tilde{x}}-\tilde{y}\hat{V}_{\tilde{y}}+\hat{V}(\tilde{x},\tilde{y}),
\end{eqnarray}

or, explicitly

\begin{equation}
\mathcal{H}_{eff}=\frac{1}{2}(H^2+D^2)\bigg(\cosh\gamma+\frac{\tilde{x}}{\sqrt{\tilde{x}^2+\tilde{y}^2}}\sinh\gamma\bigg)
\label{Ham1}
\end{equation}

As is expected, from $\gamma=0$ we recover the standard Maxwell's Hamiltonian density. Notice that the first order formalism allow us to perform the canonical analysis and obtain the effective Hamiltonian $\mathcal{H}_{eff}(\vec{D},\vec{H})$; however, it may be also convenient to view $\mathcal{H}_{eff}$ as a function of the variables $\vec{D}$ and $\vec{B}$, rather than of $\vec{D}$ and $\vec{H}$. Formally, this can be done by means of the Legendre transformation 

\begin{equation}
\tilde{\mathcal{H}}=\mathcal{H}_{eff}-\vec{H}\cdot\frac{\partial \mathcal{H}_{eff} }{\partial \vec{H}},
 \label{LG}
\end{equation}

where 

\begin{equation}
\vec{B}=- \frac{\partial \mathcal{H}_{eff} }{\partial \vec{H}}, \quad \vec{E}= \frac{\partial \mathcal{H}_{eff} }{\partial \vec{D}}.
\end{equation}

In order to implement the transformation from $\vec{H}$ to $\vec{B}$ as the fundamental variable, we need to determine the function $\vec{H}(\vec{D},\vec{B})$, which is a non trivial task due to the nonlinear constitute relations. A different approach, such as the one presented in \cite{MM1}, can be followed to obtain  $\tilde{\mathcal{H}}(\vec{D},\vec{B})$. Both Hamiltonians, $\tilde{\mathcal{H}}(\vec{D},\vec{B})$ and $\mathcal{H}_{eff}(\vec{D},\vec{H})$, are related by the Legendre transformation in (\ref{LG}).

%In Fig. \ref{Plot1} we plot the effective Hamiltonian as a function of $|\vec{H}|$ and $|\vec{D}|$ for particular values of the parameters of the model.
%\begin{center}
%\begin{figure}[h]
%\includegraphics[scale=0.5]{Heff3.pdf}
%\caption{The effective Hamiltonian $\mathcal{H}_{eff}$, with $\gamma=\frac{\pi}{4}$ and $\pi$ the angle between $\vec{H}$ and $\vec{D}$ .} \label{Plot1}
%\end{figure}
%\end{center}
To establish if the effective Hamiltonian $\mathcal{H}_{eff}$ in Eq. (\ref{Ham1}) is bounded from below we require to evaluate the limits $|\vec{H}|\rightarrow\infty$ and/or $|\vec{D}|\rightarrow\infty$. An straightforward analysis tell us that the limit  $|\vec{D}|\rightarrow\infty$ may present a negative divergency; however, in such a limit we found that $\mathcal{H}_{eff}\propto D^2 e^{-\gamma}$, which prove that the effective Hamiltonian is strictly bounded from below.

In order to look for non trivial stationary points, specifically stable minima, we must study the expressions

\begin{equation}
\frac{\partial\mathcal{H}_{eff}}{\partial D_i}=0,\quad\quad \frac{\partial\mathcal{H}_{eff}}{\partial H_i}=0. \quad i=1,2,3.
\end{equation}

which, considering $\mathcal{H}_{eff}$ as a function of $H^2$, $D^2$ and $\tilde{y}$, can be written as

\begin{equation}
2H_i\mathcal{H}_{H^2}-D_i \mathcal{H}_{\tilde{y}}=0,\quad\quad 2D_i\mathcal{H}_{D^2}-H_i \mathcal{H}_{\tilde{y}}=0,\quad i=1,2,3,
\label{Minima1}
\end{equation}

where we have defined

\begin{eqnarray}
\mathcal{H}_{H^2}&:&=\frac{\partial \mathcal{H}_{eff}}{\partial H^2}=-\frac{1}{2}(\hat{V}_{\tilde{x}}+H^2\hat{V}_{\tilde{x}\tilde{x}}+\tilde{y}\hat{V}_{\tilde{y}\tilde{x}})=\frac{1}{2}\bigg(\cosh\gamma+\frac{\sinh\gamma(H^2\tilde{y}^2+\tilde{x}^3)}{(\tilde{x}^2+\tilde{y}^2)^\frac{3}{2}}\bigg), \\ \nonumber
\mathcal{H}_{D^2}&:&=\frac{\partial \mathcal{H}_{eff}}{\partial D^2}=-\frac{1}{2}(\hat{V}_{\tilde{x}}-H^2\hat{V}_{\tilde{x}\tilde{x}}-\tilde{y}\hat{V}_{\tilde{y}\tilde{x}})=\frac{1}{2}\bigg(\cosh\gamma-\frac{\sinh\gamma((D^2-H^2)^3+8D^2\tilde{y}^2)}{(D^2-H^2)^2+4\tilde{y}^2)^\frac{3}{2}}\bigg), \\ \nonumber
\mathcal{H}_{\tilde{y}}&:&=\frac{\partial \mathcal{H}_{eff}}{\partial \tilde{y}}=-H^2\hat{V}_{\tilde{y}\tilde{x}}-\tilde{y}\hat{V}_{\tilde{y}\tilde{y}}=\frac{2\tilde{y}\sinh\gamma(D^4-H^4)}{((D^2-H^2)^2+4\tilde{y}^2)^\frac{3}{2}}.
\end{eqnarray}

Analyze the expressions in Eq. (\ref{Minima1}), for arbitrary vectors $\vec{D}=(D_x,D_y,D_z)$ and $\vec{H}=(H_x,H_y,H_z)$, is a cumbersome task; however, we can take advantage from the fact that they belong to the tensor $P^{\mu\nu}$, see Eq. (\ref{PP1}), and parametrize different configurations, all of them equivalent by Lorentz transformations. Lets notice that the vectors $\vec{D}$ and $\vec{H}$ determine a plane, which can be always be reached by adequate passive rotations of the coordinate system. We still have the freedom to perform a passive Lorentz boost in the direction perpendicular to the plane. At this step, two possibilities arise: (1) $\vec{D}$ and $\vec{H}$ are not orthogonal and can be made parallel with this boost. (2) $\vec{D}$ and $\vec{H}$ are orthogonal and will remain so after the boost. Choosing the plane of $\vec{D}$ and $\vec{H}$ as the $y-z$ plane, we realize such two cases by taking

 \begin{equation}
P_1^{\mu\nu}=\left(
\begin{matrix}
0 & 0 & 0 & -D_z\\
0 & 0 & -H_z & 0\\
0 & H_z & 0 & 0\\
D_z &0 & 0 & 0 \\
\end{matrix}
\right),\quad\quad P_2^{\mu\nu}=\left(
\begin{matrix}
0 & 0 & -D_y & 0\\
0 & 0 & -H_z & 0\\
D_y & H_z & 0 & 0\\
0 &0 & 0 & 0 \\
\end{matrix}
\right)
\end{equation}

With these parameterizations we can search for stationary points. What is found is that the expressions in Eq. (\ref{Minima1}) are satisfied only if $\cosh\gamma+\sinh\gamma=0$ or  $\cosh\gamma-\sinh\gamma=0$, which is not possible for any value of $\gamma$. The cases a) $\vec{D}\neq\vec{0}$ with $\vec{H}=\vec{0}$ and b) $\vec{D}=\vec{0}$ with $\vec{H}\neq\vec{0}$ were also investigated leading to the same conditions. We conclude that the effective Hamiltonian does not have non trivial stationary (neither minima) points. The existence of non trivial minima are relevant in different contexts, in particular within studies of spontaneous symmetry breaking or nonlinear theories, in which the linearization of the model leads to the study of linear order fluctuations on constant backgrounds. For the case of spontaneous Lorentz symmetry breaking in nonlinear electrodynamics see Refs. \cite{CP1,CP2}.  

\bigskip

The case $\vec{D}=\vec{0}$ with $\vec{H}=\vec{0}$ trivially satisfies the conditions in Eq. (\ref{Minima1}). To analyze if this stationary point always represents a local minima we can evaluate the Hessian matrix, whose rows and columns are given by the partial derivatives

\begin{equation}
M_{IJ}=:\frac{\partial^2 \mathcal{H}_{eff}}{\partial X_I \partial X_J},\quad \quad I,J=1,2,...,6. \quad \quad X_I=(D_x,D_y,D_z,H_x,H_y,H_z).
\end{equation} 

Computing the above matrix an evaluating it when $\vec{D}=\vec{0}$ and $\vec{H}=\vec{0}$ results in $M_{IJ}=\cosh\gamma\,\mathbb{1}$, where $\mathbb{1}$ represents the $6\times6$ identity matrix. We therefore confirm that this case is a local minimum of  $\mathcal{H}_{eff}$, as is expected from the limit $\gamma\rightarrow0$.

\section{Conclusions}
\label{CON1}

In this work we have studied both some aspects of the dynamics and the Hamiltonian structure of the ModMax nonlinear electrodynamics. This novel model corresponds to a nonlinear generalization of Maxwell electrodynamics that respect conformal invariance and the invariance under the $SO(2)$ duality-rotation transformations. 

Following the method of the discontinuities for nonlinear electrodynamics theories developed by Hadamard \cite{Hadamard1}, we analyzed the propagation of the discontinuities of the field. In general, the model exhibits the phenomenon of birefringence and, as discussed in the seminal ModMax article \cite{MM1}, some conditions should by imposed to avoid superluminal propagation. This behavior has been shown to appear in general models of nonlinear electrodynamics due to the possibility of the formation of caustics \cite{Boillat}. They have also been shown to arise in a study of the Euler-Heisenberg Lagrangian \cite{EH123}. Of course it is no surprise that we have encountered them in our rather different approach.

The second part of the present manuscript was devoted to the Hamiltonian analysis for which we followed  the first order formalism of Pleba\'nski for nonlinear electrodynamics models. This formalism allowed us to employ the Dirac method to identify the constraints of the model and its corresponding effective Hamiltonian. The model contains second- and first-class constraints, the latter reflecting the gauge invariance of the model. The counting of the degrees of freedom (d.o.f) reveals that the model contains two d.o.f., which satisfy the polarizations given in Eq. (\ref{ks1}), being one of them the standard Maxwell result $k^2=0$. After that we have obtained the effective Hamiltonian and proved that it is strictly bounded from below. We then investigated the possible existence of nontrivial local minima of the effective Hamiltonian. We explicitly showed that the model does not contain nontrivial minima. The existence of such nontrivial minima is relevant in nonlinear theories given that a consistent linearization of the model could be done around these points (or surfaces). 

ModMax nonlinear electrodynamics deserves further investigation, specially when is coupled to other fields. For example, for the General Relativity case, novel solutions and features are expected to appear in generalizations of black holes and gravitational waves \cite{Roman1,Roman2}. Also, standard phenomena as the Casimir effect, the Planck energy density or ultrahigh energy gamma ray propagation can be analyzed under this new model. Using current experimental data a stronger bound on the $\gamma$-ModMax parameter can be obtained from these studies. We leave these problems for future works.

\acknowledgements

C. A. E. is partially supported by the Project PAPIIT No. IN109321. The work of B. T-M. is supported by the Ph.D. scholarship program of the Universidad Aut\'onoma Metropolitana.

\end{document}